\documentclass[]{spie}  

\usepackage{amsmath,amsfonts,amssymb}
\usepackage{graphicx}
\usepackage[colorlinks=true, allcolors=blue]{hyperref}
\usepackage{enumitem}

\newcommand\micron{\mbox{$\mu$m}}

\title{The Planetary Systems Imager Adaptive Optics System: An Initial Optical Design and Performance Analysis Tools for the PSI-Red AO System}

\author[a,b]{Rebecca Jensen-Clem}
\author[b]{Philip M. Hinz}
\author[a]{M.A.M. van Kooten}
\author[c]{Michael P. Fitzgerald}
\author[d]{Steph Sallum}
\author[e]{Benjamin A. Mazin}
\author[f]{Mark Chun}
\author[a,b]{Claire Max}
\author[e]{Maxwell Millar-Blanchaer}
\author[a,b]{Andy Skemer}
\author[g]{Ji Wang}
\author[b]{R. Deno Stelter}
\author[h,i,j]{Olivier Guyon}

\affil[a]{Univ. of California, Santa Cruz (United States)}
\affil[b]{Univ. of California Observatories (United States)}
\affil[c]{Univ. of California, Los Angeles (United States)}
\affil[d]{Univ. of California, Irvine (United States)}
\affil[e]{Univ. of California, Santa Barbara (United States)}
\affil[f]{Univ. of Hawaii, Institute for Astronomy (United States)}
\affil[g]{The Ohio State Univ. (United States)}
\affil[h]{Subaru Telescope, NAOJ (United States)}
\affil[i]{The Univ. of Arizona (United States)}
\affil[j]{AstroBiology Ctr, NINS (Japan)}

\authorinfo{Further author information: (Send correspondence to R.~Jensen-Clem)\\E-mail: rjensenc@ucsc.edu}

\begin{document} 
\maketitle

\begin{abstract}
The Planetary Systems Imager (PSI) is a proposed instrument for the Thirty Meter Telescope (TMT) that provides an extreme adaptive optics (AO) correction to a multi-wavelength instrument suite optimized for high contrast science. PSI's broad range of capabilities, spanning imaging, polarimetry, integral field spectroscopy, and high resolution spectroscopy from 0.6--5\,\micron, with a potential channel at 10\,\micron, will enable breakthrough science in the areas of exoplanet formation and evolution. Here, we present a preliminary optical design and performance analysis toolset for the 2--5\,\micron\ component of the PSI AO system, which must deliver the wavefront quality necessary to support infrared high contrast science cases. PSI-AO is a two-stage system, with an initial deformable mirror and infrared wavefront sensor providing a common wavefront correction to all PSI science instruments followed by a dichroic that separates ``PSI-Red'' (2--5\,\micron) from ``PSI-Blue'' (0.5--1.8\,\micron). To meet the demands of visible-wavelength high contrast science, the PSI-Blue arm will include a second deformable mirror and a visible-wavelength wavefront sensor. In addition to an initial optical design of the PSI-Red AO system, we present a preliminary set of tools for an end-to-end AO simulation that in future work will be used to demonstrate the planet-to-star contrast ratios achievable with PSI-Red.
\end{abstract}


\section{INTRODUCTION} \label{sec:intro} 

Today's state-of-the-art high contrast imaging systems such as GPI and SPHERE have discovered and characterized approximately two dozen young, massive worlds located many astronomical units (au) from their host stars. However, unlike the 1--10 Jupiter mass mature planets at 1--10\,au that are routinely discovered by radial velocity monitoring, the occurrence rate of such young, widely separated gas giants is $<$1\%~\cite{2016PASP..128j2001B}. Characterizing the atmospheres of these abundant classes of worlds requires a 30-m telescope aperture and advanced high contrast imaging system. 

The Planetary System Imager (PSI) is a proposed second-generation instrument for the Thirty Meter Telescope (TMT). It will provide imaging, spectroscopy, and polarimetry of a diverse range of rocky planets, ice giants, and gas giants. With these unprecedented data, PSI will directly constrain the detailed physics of planetary accretion, map molecular abundances, clouds, and surface properties for rocky and giant planets, and measure the locations and compositions of thousands of planets. In addition to exoplanetary science, PSI will play an important role in fields such as Solar System science, for example by measuring the 3D structure and dynamics of planetary atmospheres at spacecraft-quality resolution ($20\,$km at Jupiter, $130\,$km at Neptune), and characterizing active processes and collisional histories in various small body populations across the Solar System.

These science topics require a broad wavelength range, spanning visible (0.6--0.9\,\micron), near/mid-IR (0.9--5.3\,\micron), and thermal IR (8--13\,\micron) wavelengths. In order to support the extreme wavefront correction required for high-contrast imaging at visible wavelengths, PSI will have a modular architecture: light from the telescope and preliminary optics will be fed to a first-stage DM and IR WFS, providing the necessary wavefront correction for light feeding instruments designed for ($>$2\,\micron). This first-stage system is called ``PSI-Red.'' Shorter wavelength light will be further directed to ``PSI-Blue,'' the second PSI module that will include a second DM and visible light WFS along with instruments optimized for 0.5--1.8\,\micron.

In this paper, we present an initial optical design of the PSI-Red AO system (Section \ref{sec:design}) and a preliminary end-to-end simulation toolset (Section \ref{sec:performance}).

\section{OPTICAL DESIGN OF PSI-RED} \label{sec:design} 

The optical design for PSI-Red is being developed to provide a common first-stage AO correction for the PSI instrument suite.  The design will include a common optical relay, a common wavefront sensor (WFS), and a deformable mirror (DM).  As part of this design effort we have worked to develop a Natural Guide Star (NGS)- based system.  This is being compared with a similar design that uses both NGS and a Laser Guide Star (LGS) mode for the TMT/MIRAO concept\cite{2006SPIE.6272E..0SC}.  The NGS system currently has the following requirements:
\begin{itemize}[noitemsep,topsep=0pt]
    \item Use over a broad wavelength range (ideally 0.5--14\,\micron),
    \item Minimum number of warm surfaces prior to PSI-Red science instrument,
    \item Minimum of 10 arcsec FOV for science instruments and WFS,
    \item 0.25\,m projected actuator spacing for the DM,
    \item WFS sampling that is about 30\% oversampled compared to the actuators, and
    \item Dichroic-based division of light.
\end{itemize}

\subsection{Layout of the Components for PSI}

PSI is intended to cover a wide wavelength range for direct characterization of exoplanets.  As such we anticipate three distinct science instruments that will be used, along with the PSI-Red WFS:

\begin{itemize}[noitemsep,topsep=0pt]
    \item PSI-Blue: 0.6--1.8\,\micron\ operation,
    \item PSI-Red: 2.0--5.1\,\micron\ operation, and
    \item PSI-10: 7.0--14\,\micron\ operation.
\end{itemize}

As a starting point for the design, we define distinct envelopes on the Nasmyth port for each of these components.   The relay is expected to occupy the center of the Nasmyth port, surrounded by general regions for the PSI-Red science instrument, PSI-Blue and the PSI-Red WFS.  
    
\subsection{PSI-Red Relay and Deformable Mirror}

The optical relay needs to reimage the f/15 TMT Nasmyth focal plane to the science instrument ports and provide a real pupil image along the way, for the placement of the DM.  The wide spectral coverage indicates the optical relay should be reflective.  Ordinarily a pair of off-axis parabolas are used for the relay.  However, a single off-axis ellipse could provide the same functionality and reduce the number of warm elements prior to entering a cold volume.

The sizing of the optical relay is determined by the choice of DM and the required actuator projected spacing of 0.25 m on the telescope entrance pupil.  Alpao Inc. currently has DMs with 64 actuators across the diameter. Based on this availability we are assuming that DMs with 128 actuators across the diameter will be manufacturable on the timescale of PSI-Red deployment.   The spacing of the actuators for these DMs is 1.5\,mm.  If we use the inner 120 actuators, this sets a pupil size of 180\,mm.  This size sets the first order parameters for our relay.  We plan for a 2:1 magnification (so that the science beam is f/30).  An off-axis ellipse with a focal length of 2,700\,mm will satisfy these constraints and create a pupil image 2.7\,m after the optic, with an f/30 focus 5.1\,m after the pupil location.  This arrangement gives us plenty of optical path to fold into areas for each of the distinct components listed above.

\begin{figure} [h]
\begin{center}
\begin{tabular}{cc}
\includegraphics[width=0.5\textwidth]{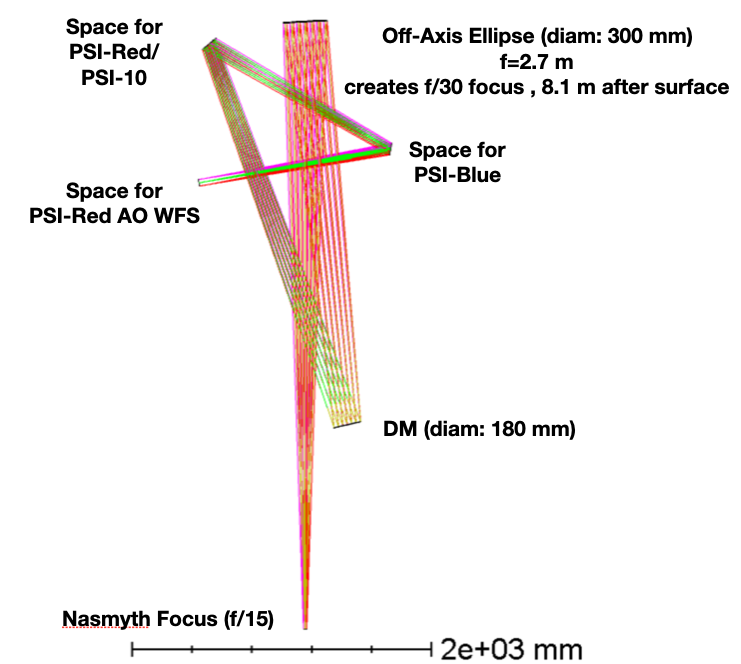}
\end{tabular}
\end{center}
\caption{Layout of the PSI-Red Optical Relay. The relay creates a pupil image at the DM with a minimum number of additional reflections.   \label{fig:relay} }
\end{figure}

After the DM, the beam is sent to the location for PSI-Red, which will be interchangeable with PSI-10.  The entrance window of the PSI-Red cryostat will be a long pass dichroic that reflects light over to the PSI-Blue portion of the layout. Interchangeable dichroics at this location will direct NIR light to the PSI-Red WFS location.  Figure~\ref{fig:relay} shows this conceptual layout.

\subsection{PSI-Red WFS}
\label{sec:psi_wfs}
\begin{figure} [h]
\begin{center}
\begin{tabular}{cc}
\includegraphics[width=0.5\textwidth]{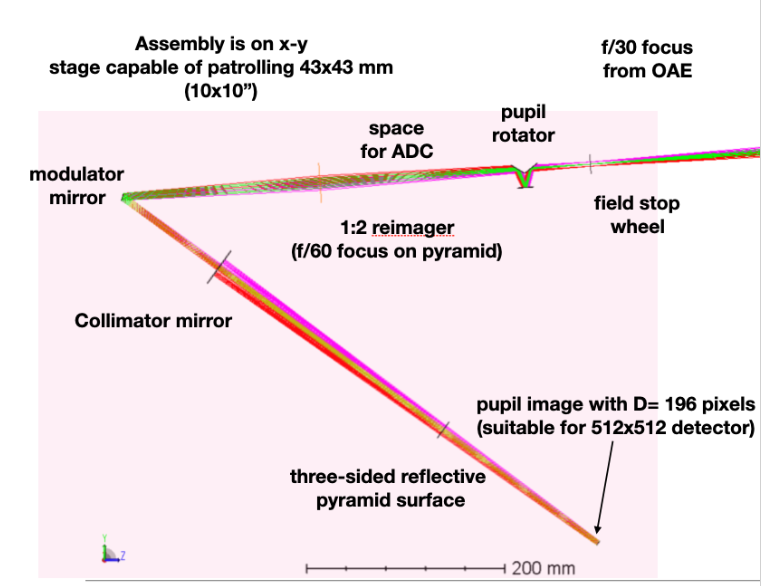} 
\end{tabular}
\end{center}
\caption{Layout of the PSI-Red WFS.  The PSI-Red relay feeds a set of optics on a patrolling x-y translation stage.  A set of reimaging optics on the board will allow for placement of mechanisms needed for pupil and field stabilization as well as image plane modulation.  These optics feed reflective pyramid wavefront sensor optics at a plate suitable for oversampling of the pupil by a SAPHIRA or similar style detector. \label{fig:wfs} }
\end{figure}

The PSI-Red WFS will use a pyramid wavefront sensor arrangement to provide aberration measurements.  The module will be located on a patrolling X-Y translation stage capable of selecting the guide star in the PSI FOV.  A 2:1 reimager will creat an f/60 image at the tip of the pyramid and provide locations for a pupil rotator, atmospheric dispersion compensator optics, and a modulator mirror for the pyramid WFS.  This design is based on a similar module developed by the Arcetri Observatory AO group for the Large Binocular Telescope\cite{2010SPIE.7736E..09E}.  Figure~\ref{fig:wfs} shows a conceptual layout of the module.   The baseline pyramid WFS optical concept is developed from a reflective three-sided design currently being tested in the UCSC Lab for Adaptive Optics\cite{2020SPIE11448E..3NS}. The pupil size will be approximately 196 pixels on the final detector. The detector is likely to be a next generation version of the SAPHIRA APD detectors with 512x512 pixels, suitable for the three pupil images from the pyramid WFS.

\subsection{PSI-Red Science Camera}

\begin{figure} [h]
\begin{center}
\begin{tabular}{cc}
\includegraphics[width=0.5\textwidth]{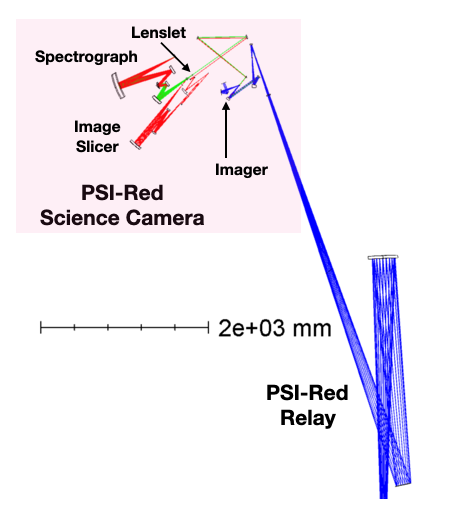} 
\end{tabular}
\end{center}
\caption{Layout of the PSI-Red Science Camera.  The camera is fed by the PSI-Red Relay.  Light entering the PSI-Red science camera can be directed to an imager or fed to a lenslet array.  The lenslet array can feed either a low resolution spectrograph directly, or, via a separate set of lenslets, an image slicer that allows for medium spectral resolution using the same spectrograph optics.  \label{fig:psiredsciencecamera} }
\end{figure}

The PSI-Red Science camera\cite{2018SPIE10702E..A5S} is based on the SCALES\cite{2020SPIE11447E..64S} design being developed for WMKO.  In brief, the design provides for an imaging camera and integral field spectrograph that are both fed via fore-optics that also provide locations for a cold stop and coronagraphic suppression. Figure~\ref{fig:psiredsciencecamera} shows an optical design of this concept, as fed by PSI-AO.  Table \ref{tab:PSIred-specs} lists the specifications of the science camera.  For the most part, the optical design does not need to change between Keck/SCALES and TMT/PSI-Red.  For diffraction-limited imaging, the field-of-view is decreased on the larger telescope, but this is acceptable for exoplanet imaging where we are primarily interested in the regions closest to the star.  However, since the exit pupil is not at infinity, and the input focal ratio is f/30 (compared to f/15 for the WMKO telescopes) the fore optics are in need of a slight redesign to be suitable for use with the PSI-Red relay.  This rework will be addressed in the conceptual development stage.

Typically, for long wavelength IR instruments that use AO, the long pass dichroic which reflects the near infrared light to PSI-Blue and the PSI-Red WFS is also the entrance window for the cryostat.  Since the dichroic is not particularly close, an evacuated tube may be needed to connect it to the PSI-Red cryostat.  Alternatively, the dichroic might be designed to have its back surface look at a ``cold sink.'' The design of this optic to minimize background will be developed in a future study.   

\begin{table}[h]
\begin{tabular}{|l|l|l|l|}
\hline
\textbf{}                                                                          & \textbf{Low-Res Spectroscopy}                                                                                                                                                                     & \textbf{Medium-Res Spectroscopy}                                                                                                           & \textbf{Imager}                                                  \\ \hline
\textbf{\begin{tabular}[c]{@{}l@{}}Wavelength/\\ Spectral Resolution\end{tabular}} & \begin{tabular}[c]{@{}l@{}}2.0-4.0$\mu$m (water ice)—R$\sim$50\\ 2.0-5.0$\mu$m (SEDs)—R$\sim$35\\ 2.9-4.15$\mu$m (L-band)—R$\sim$80\\ 3.1-3.5$\mu$m (CH4)—R$\sim$250\\ 4.5-5.2$\mu$m (M-band)—R$\sim$140\end{tabular} & \begin{tabular}[c]{@{}l@{}}2.0-2.4 (K-band)—R$\sim$4,300\\ 2.9-4.15$\mu$m (L-band)—R$\sim$2,700\\ 4.5-5.2$\mu$m (M-band)—R$\sim$6,700\end{tabular} & \begin{tabular}[c]{@{}l@{}}Filters Spanning\\ 1-5$\mu$m\end{tabular} \\ \hline
\textbf{Field-of-View}                                                             & 0.72x0.72"                                                                                                                                                                                        & 0.12x0.12"                                                                                                                                 & 6.8x6.8"                                                         \\ \hline
\textbf{Spatial Sampling}                                                          & 0.0067"                                                                                                                                                                                           & 0.0067"                                                                                                                                    & 0.0033"                                                          \\ \hline
\textbf{Coronagraphy}                                                              & \begin{tabular}[c]{@{}l@{}}Vector-Vortex + Lyot Stop\\ vAPP\\ Shaped Pupil\end{tabular}                                                                                                           & \begin{tabular}[c]{@{}l@{}}Vector-Vortex + Lyot Stop\\ vAPP\\ Shaped Pupil\end{tabular}                                                    & \begin{tabular}[c]{@{}l@{}}vAPP\\ Shaped Pupil\end{tabular}      \\ \hline
\end{tabular}
\caption{\label{tab:PSIred-specs} Top-level specifications of the PSI-Red Science Camera assuming a SCALES-like design.}
\end{table}

\subsection{A Comparison of Capabilities}

We have designed an NGS-only system in this section.  However, there may be science drivers that favor an LGS-based concept as well.  For this comparison we use the TMT/MIRAO concept\cite{2006SPIE.6272E..0SC}.

The TMT/MIRAO concept is quite similar to the design shown above.  It is intended for use with a science camera where telescope emissivity can dominate the background (and thus the noise) of an observation.  In this sense the TMT/MIRAO concept is a good comparison.  The MIRAO design is similar to that in Figure \ref{fig:relay}: it uses a reflective relay to reimage the telescope pupil onto an ambient temperature DM before relaying a 60 arcsec FOV to a science instrument.  Three LGS beacons on a radius of 70 arcsec are used for the LGS WFS.  These beams are also reimaged with the optical relay.  To do this, off-axis optics before the relay accommodate the extra back focal distance of the laser beacons. At the focal plane of the optical relay, pick off mirrors are used to redirect the LGS light to WFS cameras. This approach eliminates the use of a warm dichroic to split off the LGS light. 

The design adopted for the MIRAO relay is a pair of off-axis parabolas.  The field size, combined with the LGS reimaging drives them to have a large diameter (about 0.8 m).  An additional fold mirror is used to relay the beam to the science instrument for a total of four warm reflective optics. The PSI-Red relay adds only two warm reflective optics to the beam.  For regions of the spectrum  (3.5-4.1 $\mu$m and 10-12 $\mu$m) where telescope emissivity dominates the background, this difference can be important\cite{2020SPIE11448E..5UH}.  As an estimation of this, if we assume the primary introduces an emissivity of 4\% and the subsequent optics introduce an emissivity of 2\%, then the total emissivity of MIRAO is expected to be 16\% while the current concept will be 12\%.  For regions of the spectrum where this background dominates, this ratio defines how quickly one can reach a given signal-to-noise when comparing the two.  Thus for background-limited observations this more efficient design can reach the same SNR in only 75\% of the time. 

Table \ref{tab:comparison} summarizes the items identified in this section for comparison between the designs.

\begin{table}[h!]
\centering
\caption{A Comparison between MIRAO and the current PSI-Red AO concept}

\begin{tabular}{|c c c |} 
 \hline
 Parameter & MIRAO & PSI-Red AO \\ 
 \hline\hline
 Field of View & 60 arcsec. & 10 arcsec. \\
 \hline
Number of Warm Optics & 4 & 2 \\ 
 \hline
Expected Emissivity & 16\% & 12\% \\
 \hline
Relay Optics Diameter & 0.8 m & 0.3 m \\
 \hline
 NGS WFS Approach & NGS in instrument & Separate NIR NGS WFS module \\
 \hline
 
 LGS WFS Approach & Three beacons on 70 arcsec. radius & None \\
 \hline
 LGS Mechanism & Trombone to relocate LGS & -  \\
  & focus to input of optical relay & \\
 \hline
 Relay Focal Plane &  Accessible & In Science Instrument   \\
 \hline
 
\end{tabular}
\label{tab:comparison}
\end{table}

\newpage
\section{PERFORMANCE ANALYSIS TOOLS FOR PSI-RED} \label{sec:performance} 
The goal of the PSI-Red performance analysis simulation is to produce contrast curves that will be representative of the system's performance in order to support design, technology and science case development efforts. We have chosen to simulate each timestep in the AO system so that simulated AO telemetry, short-and-long-exposure PSFs, and short-and-long-exposure coronagraphic images can be validated against on-sky data from current high contrast imaging systems. We begin by simulating the Keck II AO system, including its near-IR pyramid wavefront sensor and $L$-band vector vortex coronagraph, and comparing the resulting simulated contrast curve with on-sky data. We then minimally update the simulation to predict the contrast achievable by the PSI-Red system. In this paper, we focus on the tools associated with the simulation itself rather than the final predicted contrast ratios for PSI-Red. 

The simulation is based on the High Contrast Imaging for Python (HCIPy\cite{por2018hcipy}) package: we make use of HCIPy's functions for creating obscured telescope apertures, atmospheric turbulence, deformable mirrors, wavefront sensors, and coronagraph optics, as well as its wavefront propagation infrastructure. Section~\ref{sim_components} below describes the major components of simulation. In~\S\ref{sec:keck_results}, we validate the results of our Keck simulation against on-sky data. In~\S\ref{sec:tmt_results}, we present our initial findings for the PSI-Red system. 

\subsection{SIMULATION COMPONENTS} \label{sim_components}

Each major component of the simulation is described below, with the parameters that differ between the Keck and TMT simulations summarized in Table \ref{tab:sim_params}.

\begin{figure} [h]
\begin{center}
\begin{tabular}{cc}
\includegraphics[width=0.4\textwidth]{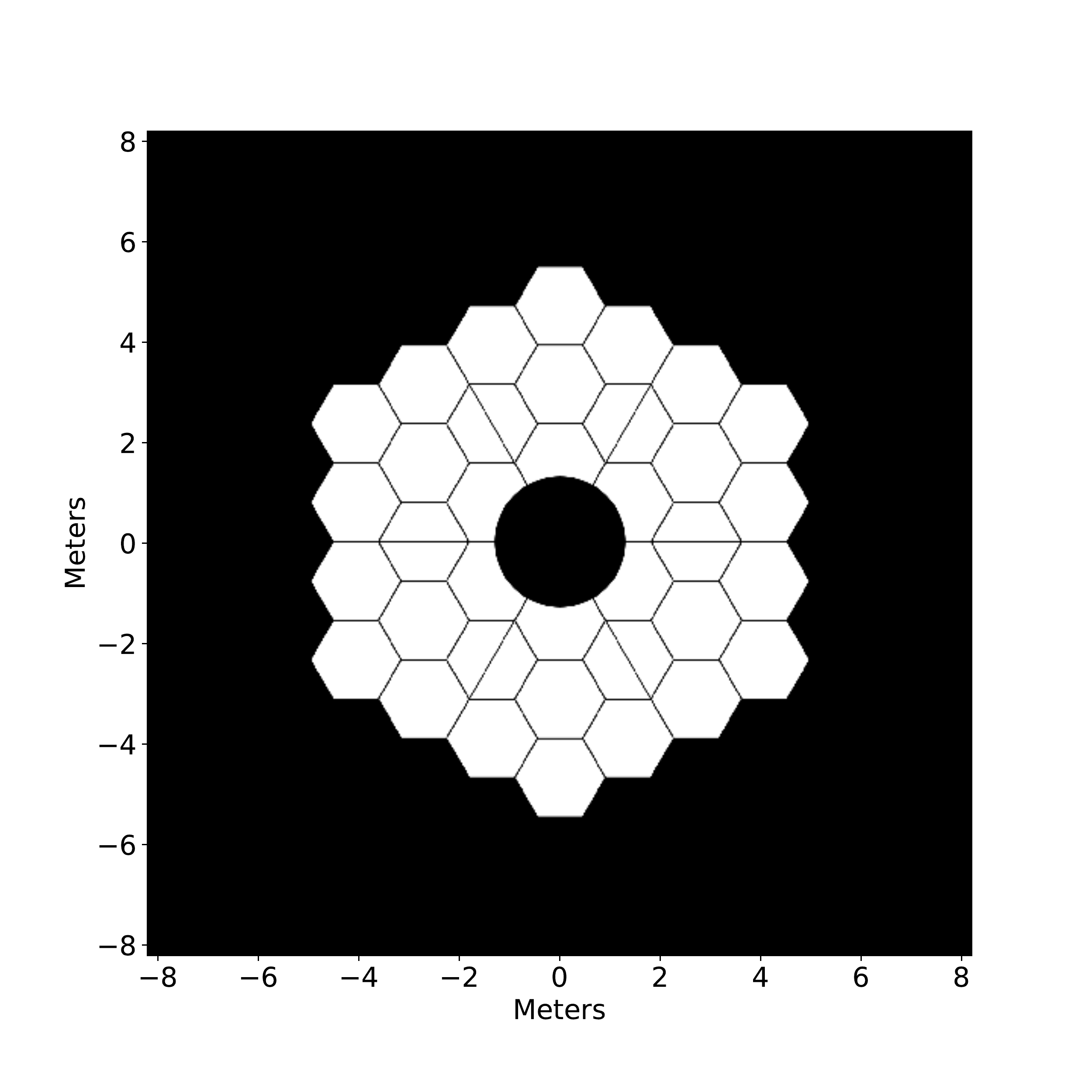} 
\end{tabular}
\end{center}
\caption{The simulated Keck aperture. Here, the aperture has been oversampled with respect to the simulations for illustration purposes (the segment gaps, spider sizes, and central obscuration remain accurate). \label{fig:keck_pupil} }
\end{figure}

\begin{figure} [h]
\begin{center}
\begin{tabular}{cc}
\includegraphics[width=0.4\textwidth]{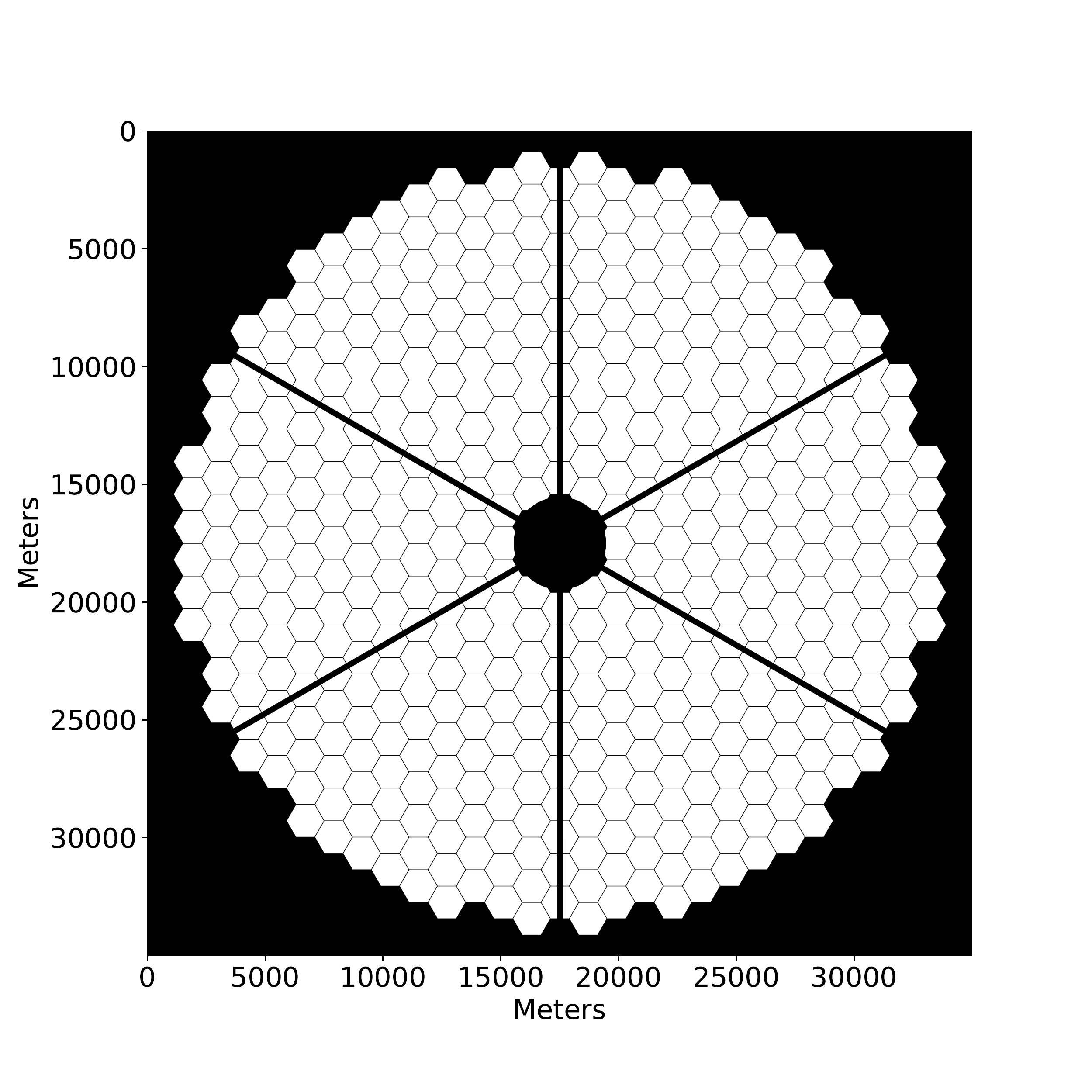} 
\end{tabular}
\end{center}
\caption{The simulated TMT aperture. Here, the aperture has been oversampled with respect to the simulations and the segment gaps have been enlarged for illustration purposes. \label{fig:tmt_pupil} }
\end{figure}

\noindent\textbf{Telescope pupil:} We begin by building the telescope pupil, including the mirror segment gaps, secondary mirror obscuration, and spiders. The Keck and TMT apertures are shown in Figures \ref{fig:keck_pupil} and \ref{fig:tmt_pupil} respectively. 

\noindent\textbf{Deformable mirror: }The DM includes 120 actuators across the pupil for the PSI-Red case and 21 actuators across the pupil for the Keck II AO case. In both cases, we assume Gaussian influence functions and infinite stroke. 

\noindent\textbf{Pyramid wavefront sensor:} We consider a modulated pyramid wavefront sensor with a modulation radius of $5 \lambda /$D. In the Keck case, we match the on-sky system's sampling of 40 pixels across the pupil. In the TMT case, we oversize the sampling by $\sim30\%$ compared with the number of DM actuators, adopting 156 pixels across the pupil and a distance of 234 pixels between pupils. Hence, a $512\times512$ pixel wavefront sensor detector would likely be sufficient for PSI-Red (see~\S\ref{sec:design}). In order to match the Keck system, we consider a slope-based reconstruction method. To this end, we cut out a circular region around each of the four pupil images on the wavefront sensor image, normalize, and compute the $x$ and $y$ slopes by the appropriate pupil image arithmetic. At each timestep of closed-loop control (described in more detail below), we sum the images from 24 sub-timesteps making up the total wavefront sensor exposure time in order to properly sample one complete modulation. We note that this simulation currently considers a four-sided pyramid wavefront sensor, whereas~\S\ref{sec:psi_wfs} considers a three-sided pyramid. This difference does not affect the analyses described here, but will be reconciled in future work. 

\noindent\textbf{Reconstruction matrix}: We first compute the interaction matrix by imposing positive and negative pokes on each DM actuator, with a probe amplitude of 2\% of the wavefront sensing wavelength of 1.65\,\micron. We then compute the reconstruction matrix by inverting the interaction matrix using HCIPy's implementation of Tikhonov regularization. 

\noindent\textbf{Atmospheric turbulence:} For the purposes of this preliminary simulation, we consider only a single layer of turbulence, assuming frozen flow. We additionally remove the tip and tilt components of the turbulence in order to avoid implementing multiple control loops with different speeds in this preliminary study. In both the Keck and TMT cases, we consider an outer scale of $80\,$m. The seeing and wind speed values for each case are given in Table~\ref{tab:sim_params}. 

\noindent\textbf{Non-atmospheric wavefront errors:} The performance of any high-contrast imaging system will be reduced by wavefront errors with a variety of spatial and temporal scales introduced by the telescope's mirrors and the AO+science system's optics. Our HCIPy simulation includes the ability to add phase errors following any peak-to-valley value and PSD power law exponent, and that refresh on any timescale smaller than the total length of the closed-loop simulation. We can also include arbitrary primary mirror segment piston errors, static or dynamic. We have not yet implemented tools for including wavefront errors due to interactions between the atmosphere and the observatory, such as dome seeing or wind loading. 

\noindent\textbf{Coronagraph: }We consider a charge-2 vector vortex coronagraph and Lyot-stop. For Keck, the Lyot stop design is based on the implemented system as described by Femen{\'\i}a Castell{\'a} et al.~2016\cite{2016SPIE.9909E..22F}, and a scaled-up version for the TMT case, further modified according to the TMT's secondary obscuration and spider placement. Optimizing the design of the PSI-Red Lyot stop will be the subject of a future study.

\noindent\textbf{Closed-loop wavefront control: }At each timestep in our closed-loop simulation, we propagate the atmospheric phase screen forward in time by 1\,ms and propagate the phase screen associated with that atmospheric turbulence through the telescope's primary mirror, DM, and pyramid wavefront sensor optics, with the optional inclusion of non-atmospheric wavefront errors. As described above, we divide the wavefront sensor's total integration time of 1\,ms into 24 subintegrations, and sum those integrates to ensure proper sampling of the modulation. We then scale the counts in the final wavefront sensor image according to the desired stellar flux and introduce photon noise. We assume that the stellar flux is the same in the wavefront sensing and science wavelenghs. We do not yet consider read noise, dark current, or sky background noise. We compute the slopes as described above, subtract their mean, and update the DM actuator positions as follows: 
\begin{equation}
    A_{t+1} = A_t - g\left ( RM \times s_t \right ) 
\end{equation}
where $A_{t+1}$ are the updated DM actuator positions, $A_t$ are the current actuator positions, $g$ is the gain (we select $g=0.4$), $RM$ is the reconstruction matrix, and $s_t$ are the just-measured slopes. Before continuing to the next timestep, we compute the Strehl via a non-coronagrpahic PSF image, and create our science image by propagating the wavefront through the coronagraph optics. 

\begin{table}[h!]
\centering
\caption{The simulation parameters that differ between the Keck and TMT cases}

\begin{tabular}{|c c c |} 
 \hline
 Parameter & Keck value & PSI-Red value \\ 
 \hline\hline
DM actuators across the pupil & 21 & 120 \\ 
 \hline
PYWFS detector pixels across each pupil & 40 & 156 \\
 \hline
Seeing (500\,nm) &  $0.66^{\prime\prime}$ & $0.5^{\prime\prime}$ \\
 \hline
Wind speed & 8.8\,m/s & 10\,m/s \\
\hline
Guide star magnitude & $H$=4 & $H$=5 \\
\hline
Science wavelength & 3.776\,\micron & 2.2\,\micron \\
 \hline 
\end{tabular}
\label{tab:sim_params}
\end{table}

\subsection{SIMULATION VALIDATION USING KECK DATA} \label{sec:keck_results}

\begin{figure} [h]
\begin{center}
\begin{tabular}{cc}
\includegraphics[width=\textwidth]{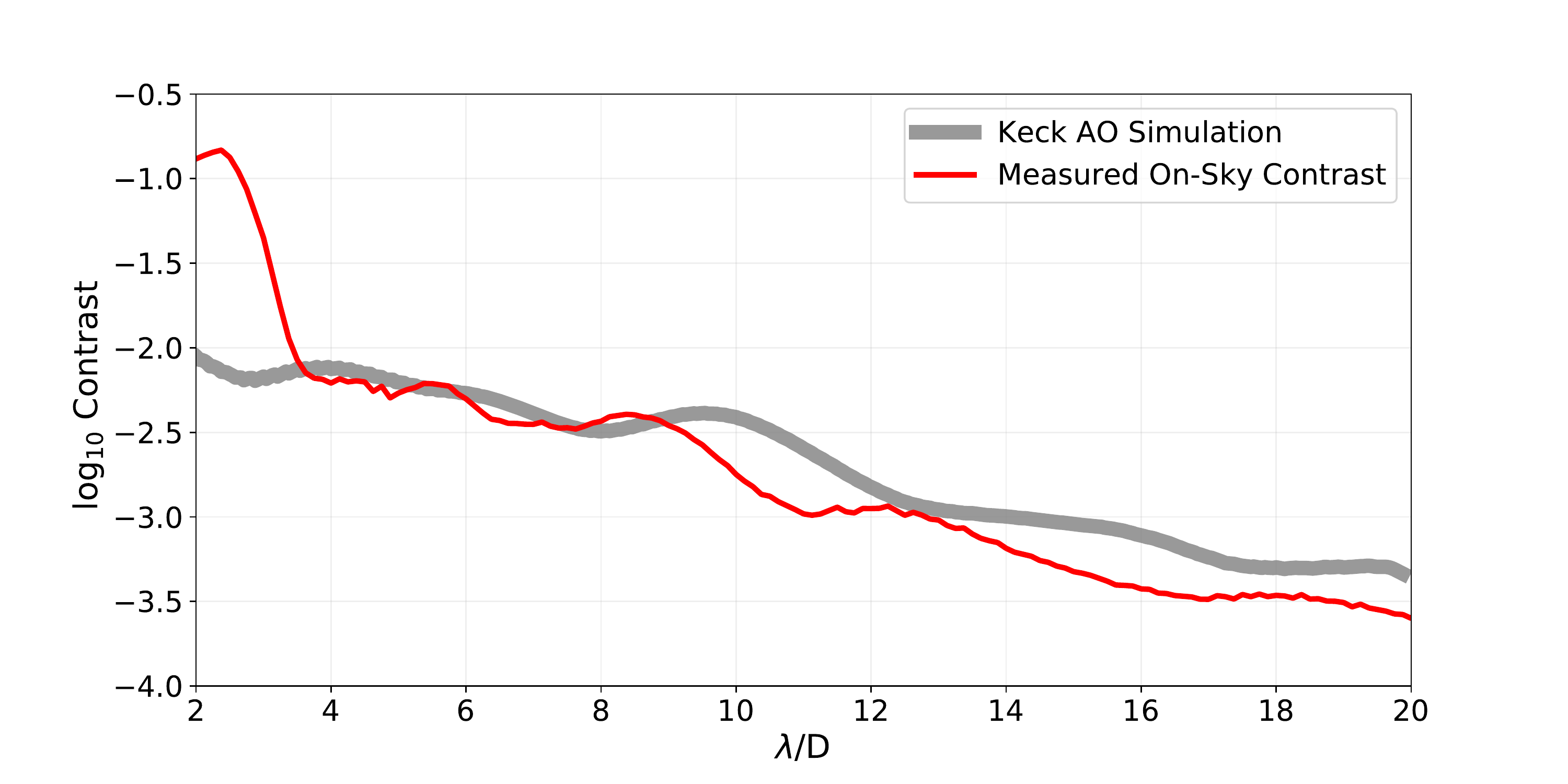} 
\end{tabular}
\end{center}
\caption{A comparison of the median of thirty-three 18-s unprocessed contrast curves on-sky at Keck (red) and simulated here (gray). The two curves deviate at small separations due to the tip/tilt errors that are not captured in our simulation.\label{fig:keck_noise_comparison} }
\end{figure}

In order to validate the results of our simulation, we consider a Keck II observation of Theta Hya ($H$=4) on December 26th 2020. During these $L$-band NIRC2 observations, we used Keck's $H$-band pyramid wavefront sensor\cite{2020JATIS...6c9003B} and the $L$-band vector vortex coronagraph. We used the automated pipeline described in Xuan et al.~2018\cite{2018AJ....156..156X} to bad pixel correct, flat-field correct, sky correct, and register the NIRC2 images. Each NIRC2 image represented a total exposure time of $18\,$s. For each image, we used the Vortex Image Processing Package (VIP\cite{2017AJ....154....7G}) to compute the student-$t$ corrected per-frame contrast curve (because we have not used ADI, PCA, etc, this can be considered a ``raw'' or unprocessed contrast curve). The median of 33 of these contrast curves is shown in Figure \ref{fig:keck_noise_comparison}.

We used data from the Mauna Kea Weather Center to compute the average wind speed (8.8\,m/s) and seeing ($0.66^{\prime\prime}$) during this observation, and include those values in the atmospheric turbulence component of our simulation. We further adopted Theta Hya's magnitude of $H$=4. 

\begin{figure} [h]
\begin{center}
\begin{tabular}{cc}
\includegraphics[width=\textwidth]{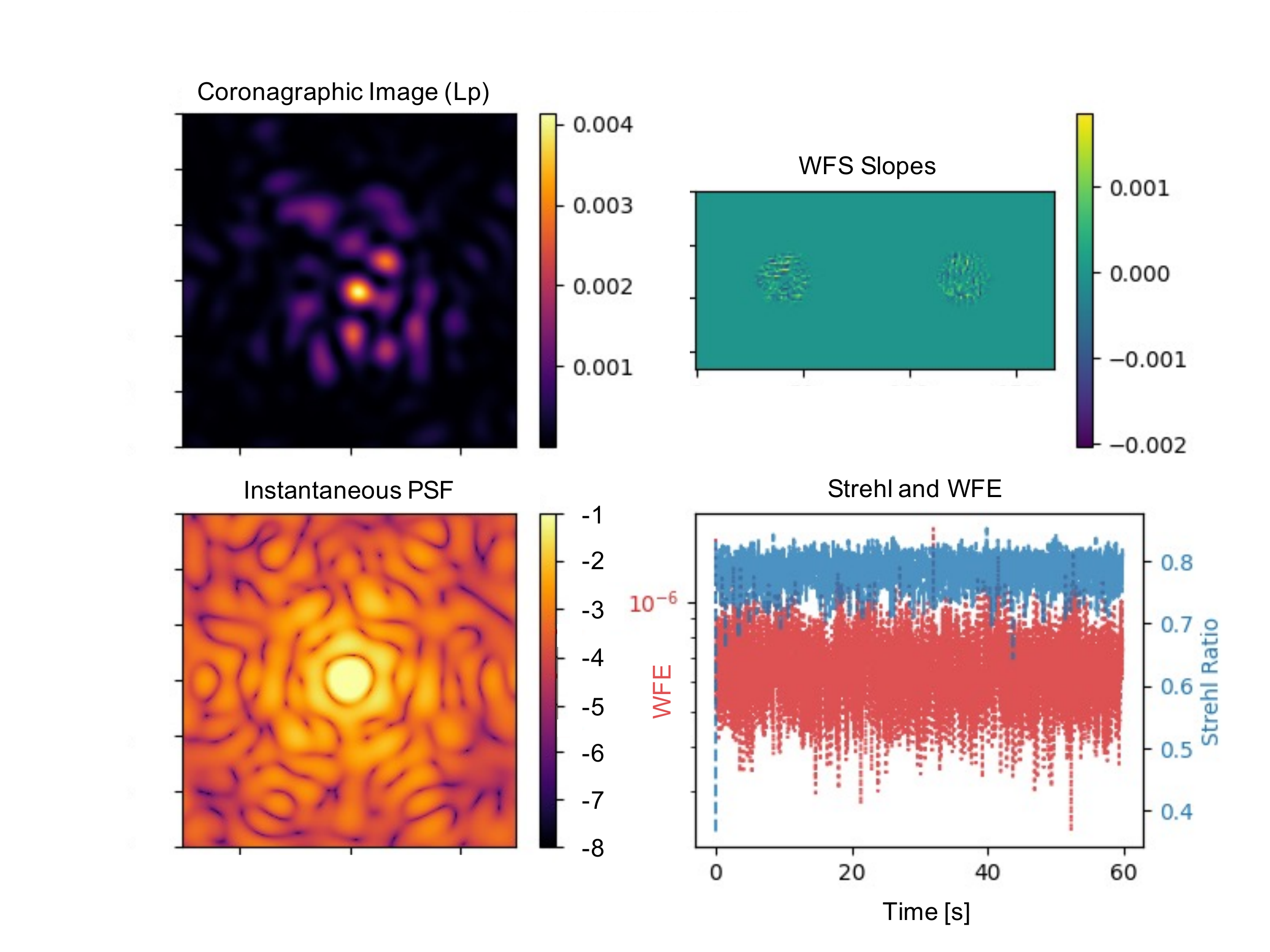}
\end{tabular}
\end{center}
\caption{Sample output from our Keck II AO Simulation. Upper left: an instantaneous $L$-band coronagraphic image. Upper right: the wavefront slopes measured from the pyramid wavefront sensor. Lower left: an instantaneous $H$-band PSF. Lower right: the Strehl ratio and total WFE versus simulated time. \label{fig:keck_sim} }
\end{figure}

More difficult are decisions regarding non-atmospheric wavefront errors: ideally, we would introduce primary mirror segment piston errors and wavefront errors before and after the wavefront sensor with a range of timescales, as indicated by measurements from the Keck AO bench and NIRC2 system. However, such errors have not yet been characterized at the level of detail that would be required to include them in this simulation. Rather than choosing arbitrary or degenerate WFE locations, amplitudes, and timescales, we elect to consider static errors from the primary mirror segment pistons only for the purposes of this preliminary study. We accomplish this by drawing random piston offsets from a normal distribution in order to ensure a particular RMS WFE and tuning that WFE such the final contrast curve most closely matched the on-sky data. We find that a WFE of 120\,nm RMS produced the best match -- this is similar to the total estimated wavefront error of $230\,$nm (when represented on the primary mirror, this is $230/2=115\,$nm) for bright NGS targets at Keck II AO. We emphasize that any dependence of the residual WFE on temporal effects is conditional on the accuracy of the above assumptions. Figure \ref{fig:keck_sim} shows examples of the instantaneous coronagraphic image, PSF, and slopes measured by the pyramid wavefront sensor for this simulation, as well as the Strehl ratio and WFE versus time. 

For each second of simulation time, we saved the PSF and coronagraphic image. In order to match the on-sky exposures, we then summed these data to create thirty-three 18-s exposure PSF frames and coronographic frames. We then used VIP to compute the student-$t$ corrected per-frame contrast curve. The median of the resulting thirty-three noise curves is shown in Figure \ref{fig:keck_noise_comparison}. 

\subsection{PSI-RED DEMONSTRATION} \label{sec:tmt_results}

Having benchmarked our Keck simulation against on-sky contrast curves, we now seek to make the smallest number of changes to the Keck simulation to represent TMT/PSI-Red. We start by simulating the TMT aperture (Figure \ref{fig:tmt_pupil}) and updating the sampling associated with the DM and the PYWFS. A summary of the differences between the Keck and PSI-Red simulations are listed in Table \ref{tab:sim_params}. 

\begin{figure} [h]
\begin{center}
\begin{tabular}{cc}
\includegraphics[width=\textwidth]{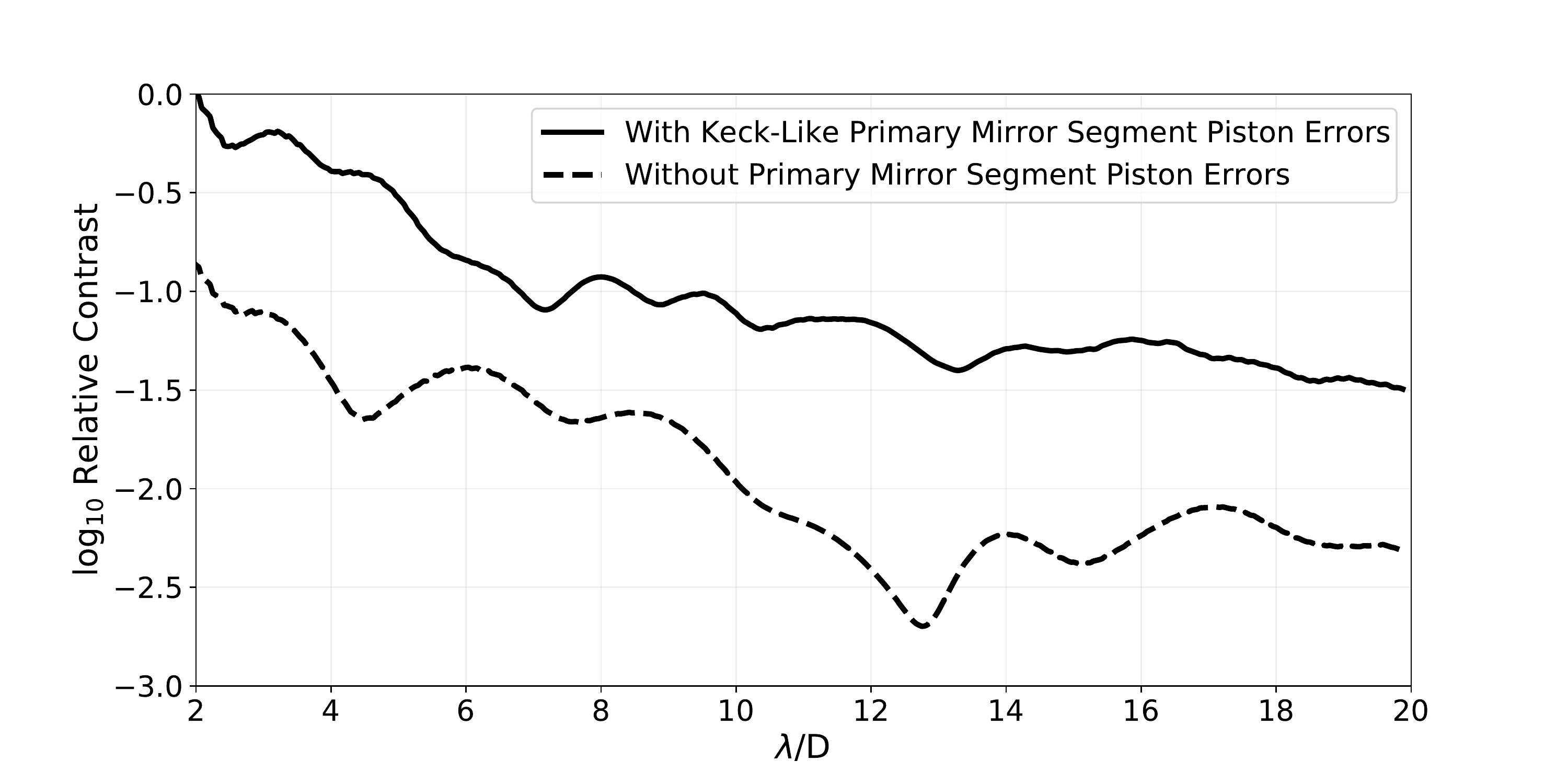} 
\end{tabular}
\end{center}
\caption{A comparison of simulated TMT relative contrast curves given the same primary mirror segment piston RMS WFE as in our Keck simulation (solid line) and without any piston errors (dashed line). We note that the contrast has been normalized to the contrast of the Keck-like primary mirror segment piston error case at $2\lambda/$D to highlight the difference between these curves. \label{fig:tmt_contrast} }
\end{figure}

We simulated 30-s total of $K$-band coronagraphic science images. Figure \ref{fig:tmt_contrast} shows the results: when WFEs that were required to match the Keck on-sky contrast are introduced, the contrast suffers by a factor of a few. We note that the contrast has been normalized to the contrast of the Keck-like primary mirror segment piston error case at $2\lambda/$D to emphasize the difference between these curves rather than their absolute contrasts. 

These results highlight the urgent need to characterize the AO and contrast error budgets at today's state-of-the-art high contrast imaging facilities. Simulations that seek to predict the performance of future high contrast instruments such as PSI must be validated against on-sky data, and hence today's instruments have an important role to play in risk reduction for future facilities. 

In summary, improving the fidelity of this simulation will require a range of future steps:
\begin{itemize}[noitemsep,topsep=0pt]
    \item Estimate the residual primary mirror segment piston and tip/tilt errors at Keck II given active measurements from the recently installed Zernike wavefront sensor during science operations and include them in this simulation (simulation infrastructure is in place).
    \item Estimate the wavefront aberrations introduced by optics with particular surface quality estimates in current systems and include them in this simulation (simulation infrastructure is in place).
    \item Estimate the time-evolving wavefront aberrations in current systems and include them in this simulation (simulation infrastructure is in place).
    \item Include a high-speed tip/tilt loop in the PSI-Red AO simulation.
    \item Design an optimal Lyot stop for the TMT aperture.
    \item Include advanced wavefront control methods such as predictive control (simulation infrastructure is in place) and focal plane wavefront sensing. 
\end{itemize}

As the above steps are completed, we will expand the scope of the simulations by exploring the post-processed contrast as a function of guide star magnitude, AO loop speed, and parallactic rotation.

\clearpage



\acknowledgments 
This work was partially supported by the National Science Foundation AST-ATI Grant 2008822, the Heising-Simons Foundation, and the University of California Observatories minigrant program. RJ-C also thanks UC Santa Cruz Profs.~Brant Robertson and Daniel Fremont for computational resources used as part of this study. The authors thank all participants in the PSI collaboration and all staff members at the TMT and WMKO for their support of this effort. 
  


\bibliography{report} 
\bibliographystyle{spiebib} 

\end{document}